\documentclass{amsart}

\title{Sampling theorem based Fourier-Legendre transform}

\author{S. Kuwata${}^1$, K. Kawaguchi${}^2$}

\address{${}^1$Graduate School of Information Sciences, Hiroshima City University,  Asaminami-ku, Hiroshima 731-3194, Japan}

\address{${}^2$Department of System Engineering, Hiroshima City University, Asaminami-ku, Hiroshima 731-3194, Japan}


\usepackage{mathptmx} 

\newtheorem{proposition}{Proposition}
\newtheorem{remark}{Remark}

\begin{document}
\maketitle

\begin{abstract}

The product of any number of Legendre functions, under a restricted domain, can be expanded by
the corresponding Legendre polynomials, with the coefficient being the sinc function.
While an analogous expansion can be made for any number of Gengenbauer functions,
it is not allowed for more than two Jacobi functions.
To obtain such an expansion, the sampling theorem is of great availability.

\end{abstract}

\section{Introduction}
\label{sec:intro}

The Legendre function $P_\nu (x)$ and its relatives are a powerful tool in analyzing various potential problems~\cite{addition}.
One of the specific properties of $P_\nu (x)$ (for $- 1 < x \leq 1$) is that its Fourier transform with respect to $\nu$ has a compact support lying within the interval $[ - \pi, \pi ]$.
This implies that by the sampling theorem, $P_\nu$ can be expanded with respect to the Legendre polynomial $P_n$, with the expansion coefficient being the sinc function.
An analogous expansion holds in the case where $P_\nu$ is replaced by the Jacobi function $P_\nu^{(\alpha, \beta)}$~\cite{jacobi2,jacobi1}.
However, it is not certain that there may exist an analogous expansion for the product $P_\nu^{(\alpha, \beta)} (x) \, P_\nu^{(\alpha', \beta')} (y)$;
it is only known that such an expansion can be realized in the case of $\alpha = \beta = \alpha' = \beta'$~\cite{gradshteyn}, in which
$P_\nu^{(\alpha, \beta)}$ is reduced to the Gengenbauer function $C_\nu^\gamma$ ($2\gamma = \alpha + \beta + 1$).

The aim of this paper is to examine whether or not the sampling theorem based summation formula holds for the product of $N$ Jacobi functions $P_\nu^{(\alpha, \beta)}$'s ($N \in \mathbb N$),
to reveal that while $N$ can be chosen as arbitrary as long as $\alpha = \beta$, the case of $N \geq 3$ is not allowed unless $\alpha = \beta$.
In Sec.~\ref{sec:legendre}, we deal with the sampling theorem based summation formula for the product of multiple $P_\nu$'s,
together with the calculation of an infinite series $G$ which is vanishing for a domain where the sampling theorem can be applied.
This kind of calculation turns out to be useful in obtaining an analogous summation formula where $P_\nu$ is replaced by $C_\nu^\gamma$.
In Sec.~\ref{sec:jacobi},
we obtain the sampling theorem based summation formula for all the allowed products of $N$ Jacobi functions $P_\nu^{(\alpha, \beta)}$'s.
%
Conclusion is given in Sec.~\ref{sec:conclusion}, where the Hermite function $H_\nu$ is dealt with.

\section{Fourier-Legendre transform}
\label{sec:legendre}

To begin with, it is useful for the present purpose to represent the Legendre function $P_\nu (x)$ (for $\nu \in \mathbb C)$ using the Mehler-Dirichlet integral~\cite{whittaker}
\begin{align}
P_\nu (\cos \theta) = \frac{\sqrt{2}}{\pi} \int_{0}^\theta \frac{\cos \left[ (\nu + \frac{1}{2}) \psi \right] }{\sqrt{\cos \psi - \cos \theta}} \, {\rm d} \psi \qquad (0 < \theta < \pi),
\label{eq:mehler}
\end{align}
which can be applied to $\theta = 0$ by taking a limit of $\theta \rightarrow 0$ in the right-hand side.
From eq.~(\ref{eq:mehler}), it is found that the support of the Fourier transform of $P_\nu$ with respect to $\nu$ lies within the interval $[ - \pi, \, \pi ]$.
This implied that by the sampling theorem, $P_\nu (\cos \theta)$ can be expanded with respect to $P_n (\cos \theta)$ ($n \in \mathbb Z$),
with the expansion coefficient being the sinc function,
provided that $0 \leq \theta < \pi$.

For the product of multiple $P_\nu$'s, the analogous expansion can be made, with the result that for $f_\nu^{(N)} := \prod_{i=0}^N P_\nu (\cos \theta_i)$
\begin{align}
f_\nu^{(N)} = \sum_{n=-\infty}^\infty {\rm sinc} \, (\nu - n) \, f_n^{(N)} \qquad \left( \sum_{i=1}^N | \theta_i | < \pi \right),
\label{eq:f_nu}
\end{align}
where ${\rm sinc} \, z = \frac{\sin \pi z}{\pi z}$.
Unless $\sum_{i=1}^N | \theta_i | < \pi$, it seems somewhat more complicated to expand $f_\nu^{(N)}$ using $f_n^{(N)}$'s ($n \in \mathbb Z$).
Here we instead calculate a simplified series such that is identically vanishing for $\sum_{i=1}^N | \theta_i | < \pi$.
Such a series can be realized by recalling that
$P_\nu (x)$ satisfies the second order differential equation ${\mathcal H}_\nu P_\nu (x) = 0$, where
${\mathcal H}_\nu = (1-x^2) \frac{{\rm d}^2}{{\rm d}x^2} - 2 x \frac{\rm d}{{\rm d} x} + \nu (\nu+1)$.
Applying ${H}_\nu$ to both sides of eq.~(\ref{eq:f_nu}) and using $P_{- n-1} (x) = P_n (x)$ for all $n \in \mathbb Z$, we obtain
\begin{align*}
G = 0 \qquad  \left( \sum_{i=1}^N | \theta_i | < \pi \right),
\end{align*} 
where $G = G(\cos \theta_1, \ldots, \cos \theta_N) =  \sum_{n=0}^\infty (-1)^n \, (2n+1) \, f_n^{(N)}$.
%
%
%
The calculation of $G$ has already been made up to $N=3$, so we concentrate on the case of $N=4$.
However, we review the case of $N \leq 3$, for completeness.%

This kind of calculation turns out to be useful in generalizing eq.~(\ref{eq:f_nu}) where $P_\nu$ is replaced by $C_\nu^\gamma$,
as will be discussed in the latter half of Sec.~\ref{sec:jacobi}.

\subsection{$N=1$}
\label{subsec:N=1}

In this case, $G$ is closely related to the generating function of $P_n (\cos \theta)$ as
\begin{align*}
\sum_{n=0}^\infty t^n  P_n (\cos \theta) = \frac{1}{\sqrt{1 - 2 t \cos \theta + t^2}} \qquad \left (| t | < 1 \right),
\end{align*}
from which we obtain $g_t (x) := \sum_{n=0}^\infty  (2n+1) \, t^n \, P_n (x) = \frac{1-t^2}{(1-2 t x + t^2)^{\frac{3}{2}}}$ for $| x | \leq 1$.
Considering that $g_t (x)$ has the following properties
\begin{align*}
\begin{cases}
\lim_{t \searrow -1} \, g_t (x) = 0 & (x \neq -1), \\
\int_{-1}^1 g_t (x) \, {\rm d} x = 2 & (t \in (-1, \, 1) )
\end{cases}
\end{align*}
we can regard $\lim_{t \searrow -1} g_t (x)$ as $2 \, \delta (x+1)$ for $| x | \leq 1$,
where $\delta (x)$ represents the Dirac delta.
Hence, we obtain
\begin{align}
G (\cos \theta) = 2 \, \delta (\cos \theta + 1).
\label{eq:G1}
\end{align}
Notice that eq.~(\ref{eq:G1}) can be derived more easily from eq.~(\ref{eq:G2}) below, by using $P_n (1) = 1$ for all $n \in \mathbb N$.

\subsection{$N=2$}
\label{subsec:N=2}

In this case, the convenient way to obtain $G$ is to use the completeness relation for $P_n (x)$ as
%
\begin{align*}
\sum_{n=0}^\infty (2n + 1) \, P_n (x) \, P_n (y) = 2 \, \delta (x-y) \qquad (- 1 \leq x,y \leq 1).
\end{align*}
Recalling that $P_n (-y) = (-1)^n \, P_n (y)$ for $n \in \mathbb N$, we find that
\begin{align}
G (\cos \theta_1, \cos \theta_2) &= 2 \, \delta (\cos \theta_1 + \cos \theta_2) \notag \\
& = \delta \left( \cos \frac{\theta_1 + \theta_2}{2} \cdot \cos \frac{\theta_1 - \theta_2}{2} \right), 
\label{eq:G2}
\end{align}
which confirms that $G$ turns out to be vanishing for $\sum_{i=1}^2 | \theta_i | < \pi$, where the sampling theorem can be applied.

\subsection{$N=3$}
\label{subsec:N=3}

In Ref.~\cite{formula,watson}, $G$ is calculated as
\begin{align}
G(\cos \theta_1, \cos \theta_2, \cos \theta_3) =
\begin{cases}
0 & (\eta_3 > 0), \\
\frac{2}{\pi} \frac{1}{\sqrt{- \eta_3}} & (\eta_3 < 0),
\end{cases}
\label{eq:G3}
\end{align}
where $\eta_3$ is given by
\begin{align*}
\eta_3 & =
2^2 \cos \frac{\theta_1 + \theta_2 + \theta_3}{2} \cdot \cos \frac{- \theta_1 + \theta_2 + \theta_3}{2}
\cdot \cos \frac{\theta_1 - \theta_2 + \theta_3}{2} \cdot \cos \frac{\theta_1 + \theta_2 - \theta_3}{2} \\
&=
\sum_{i=1}^3 \cos^2 \theta_i + 2 \prod_{i=1}^3 \cos \theta_i - 1.
\end{align*}
Notice that $\eta_3 > 0$ for $\sum_{i=1}^3 | \theta_i | < \pi$, in which $G$ turns out to be vanishing.

For the sake of completeness, we will derive the relation of (\ref{eq:G3}). A useful relation is the one derived from the addition theorem for $P_n (x)$ as~\cite{addition}
\begin{align}
P_n (\cos \alpha) \, P_n (\cos \beta) = \frac{1}{\pi} \int_0^\pi P_n (\cos \Omega) \, {\rm d} \omega,
\label{eq:addition}
\end{align}
where $\cos \Omega = \cos \alpha \, \cos \beta + \sin \alpha \,  \sin \beta \, \cos \omega$.
Substituting eq.~(\ref{eq:addition}) into the definition of $G$ for $N=3$ and using eq.~(\ref{eq:G2}), we can obtain eq.~(\ref{eq:G3}) quite straightforwardly;
whereas eq.~(\ref{eq:G3}) reduces to eq.~(\ref{eq:G2}) by taking the limit of $\theta_3 \rightarrow 0$.


\subsection{$N=4$}
\label{subsec:N=4}

Before proceeding further, we generalize $\eta_3$ for the present purpose.
Let $\eta_4 = \eta_4^{\, +}, \eta_4^{\, -}$ be defined as
\begin{align*}
\eta_4 = 2^2 \prod_{i=1}^4 \cos \Phi_i, \qquad (\Phi_1, \Phi_2, \Phi_3, \Phi_4) = (\theta_1, \theta_2, \theta_3, \theta_4) \, U,
\end{align*}
where $U = U_+, U_-$ (for $\eta_4 = \eta_4^{\, +}, \eta_4^{\, -}$, respectively) can be written using a hermitian unitary matrix as
\begin{align*}
U_+ = - \frac{1}{2} \left( \begin{smallmatrix} 1 & 1 & 1 & -1 \\ 1 & 1 & -1 & 1 \\ 1 & - 1 & 1 & 1 \\ -1 & 1 & 1 & 1 \end{smallmatrix} \right) , \qquad
U_- = - \frac{1}{2} \left( \begin{smallmatrix} 1 & -1 & -1 & 1 \\ -1 & 1 & -1 & 1 \\ -1 & - 1 & 1 & 1 \\ 1 & 1 & 1 & 1 \end{smallmatrix} \right).
\end{align*}
(It may be intriguing to point out that $U_+$ coincides with the modular $S$ matrix acting on the characters of four admissible representation of
$\widehat{\mathfrak su(3)}_{-\frac{3}{2}}$~\cite{buican}.)
It should be remarked that for all $i, j \in \{ 1,2,3,4 \}$, we have
\begin{align*}
\begin{cases}
\eta_4^{\, \pm} \rightarrow \eta_4^{\, \pm} & (\theta_i \leftrightarrow \theta_j), \\
\eta_4^{\, +} \leftrightarrow \eta_4^{\, -} & (\theta_i \rightarrow - \theta_i), \\
\eta_4^{\, \pm} \rightarrow \eta_3 & (\theta_4 \rightarrow 0).
\end{cases}
\end{align*}
As in the case of $\eta_3$, it is to be noted that $\eta_4^{\, \pm} > 0$ for $\sum_{i=1}^4 | \theta_i | < \pi$.

Now we are in a position to calculate $G$ for $N=4$.
As in the case of $N=3$, substituting eq.~(\ref{eq:addition}) into the definition of $G$ for $N=4$ and using eq.~(\ref{eq:G3}), we obtain
\begin{align*}
G(\cos \theta_1, \ldots, \cos \theta_4) &= \frac{2}{\pi^2} \int_0^\pi \frac{1}{\sqrt{ -\eta} } \, \Theta (-\eta) \, {\rm d} \omega,
\end{align*}
where $\Theta (x) = 1$ (for $x \geq 0$), $\Theta (x) = 0$ (for $x < 0$); and
$- \eta = 1 - \cos^2 \theta_3 - \cos^2 \theta_4 - 2  \cos \theta_3 \, \cos \theta_4 \, \cos \Omega - \cos^2 \Omega $, with
$\cos \Omega = \cos \theta_1 \, \cos \theta_2 + \sin \theta_1 \, \sin \theta_2 \, \cos \omega$.
By changing the variable $\omega \in [0, \, \pi]$ to $t \in [0, \, \infty)$ as $t = \tan \frac{\omega}{2}$, $G$ can be rewritten as
\begin{align*}
G (\cos \theta_1, \ldots, \cos \theta_4) = \left( \frac{2}{\pi} \right)^2 \int_{\mathbb R_+ \cap \, I}  \frac{1}{\sqrt{A t^4 + B t^2 + C}} \, {\rm d} t,
\end{align*}
where $\mathbb R_+ = \{ x \in \mathbb R \, | \, x \geq 0 \}$ and
$I= \{ x \in \mathbb R \, | \, A x^4 + B x^2 + C \geq 0 \}$. Here,
$B$ and $D \; (= B^2 - 4AC)$ are related to $\eta_4^{\, \pm}$ through the following relations:
\begin{align*}
\begin{cases}
B = - (\eta_4^{\, +} + \eta_4^{\, -} ), \\
D = ( \eta_4^{\, +} - \eta_4^{\, -} )^2 \; (\geq 0).
\end{cases}
\end{align*}
Noticing further that $B$ is related to $A, C$ as
\begin{align*}
B &= A + C + 4 \sin^2 \theta_1 \, \sin^2 \theta_2 \\
& \geq A + C,
\end{align*}
we find that $( A, C > 0) \Longrightarrow (B > 0)$.
Moreover, change the variable $\mathbb R_+ \ni t \mapsto x \in \mathbb R_+$ as $x = t^2$. Then $G$ can be replaced by
\begin{align*}
G (\cos \theta_1, \ldots, \cos \theta_4) = \frac{2}{\pi^2} \int_{\mathbb R_+ \cap \, J}  \frac{1}{\sqrt{x}} \frac{1}{\sqrt{Ax^2 + Bx + C}} \, {\rm d} x,
\end{align*}
where $J = \{ x \in \mathbb R \, | \, A x^2 + Bx + C \geq 0 \}$.
The domain of integration, $\mathbb R_+ \cap J$, depends on the sign of $A$, $B$, and $C$.
The results are summarized in Table~\ref{t:class}, where $a$ and $b$ ($a \leq b$) represent the root of $Ax^2 + Bx + C = 0$,
so that $\{ a, b \} = \{ \frac{\eta_4^{\, +}}{A}, \frac{\eta_4^{\, -}}{A} \}$.

\begin{table}[bth]
\caption{Domain of integration $\mathbb R_+ \cap J$, where $a$ and $b$ are chosen as $a \leq b$.
Recall that the case of $(A, B,C ) = (+,-,+)$ is not allowed, due to the condition of $B \geq A + C$.}
\centering
\begin{tabular}{cccccc}
\hline\noalign{\smallskip}
$A$ & $B$ & $C$ & $\mathbb R_+ \cap J$ & $(\eta_4^{\, +}, \eta_4^{\, -})$ & $G$ \\
\noalign{\smallskip}\hline\noalign{\smallskip}
$-$ & $-$ & $-$ & $\emptyset$ & $(+, +)$ & $0$ \\
$-$ & $-$ & $+$ & $[0, \, b]$ & $(-, +)$ & $S_3$ \\
$-$ & $+$ & $-$ & $[a, \, b]$ & $(-, -)$ & $S_1$ \\
$-$ & $+$ & $+$ & $[0, \, b]$ & $(-,+)$ & $S_3$ \\
$+$ & $+$ & $-$ & $[b,  \infty)$ & $(+,-)$ & $S_2$ \\
$+$ & $+$ & $+$ & $[0,  \infty)$ & $(-, -) $ & $S_1$ \\
$+$ & $-$ & $-$ & $[b, \infty)$ & $(+,-)$ & $S_2$ \\
$+$ & $-$ & $+$ & (NA) & (NA) & (NA) \\
\noalign{\smallskip}\hline
\end{tabular}
\label{t:class}
\end{table}

The rest is to perform the integration with respect to $x$ over $\mathbb R_+ \cap J$, with the result given by the complete elliptic integral
${\boldsymbol K} (k) = \frac{\pi}{2} \, {}_2 F_1 \left( \frac{1}{2}, \frac{1}{2}; 1; k^2 \right)$.
To begin with,
consider the case of $(A,B,C) = (-, + , -)$, in which $(\eta_4^{\, +}, \eta_4^{\, -}) = (-,-)$.
In this case, we denote $G$ by $S_1$, where $S_1$ can be written as
$S_1 = \frac{2}{\pi^2} \frac{1}{\sqrt{-A}} I (a,b)$, with 
$I (a,b) = \int_a^b \frac{1}{\sqrt{x}} \frac{1}{\sqrt{x-a}} \frac{1}{\sqrt{b-x}}  \, {\rm d} x$.
By the formula~\cite{formula}, we have
\begin{align*}
I (a,b) =
\begin{cases}
\pi \frac{1}{\sqrt{a}} \, {}_2 F_1 \left ( \frac{1}{2}, \frac{1}{2}; 1 ; \frac{a-b}{a} \right)
& \left( \left| {\rm arg} \left( \frac{b}{a} \right) \right| < \pi  \right), \\
\pi \frac{1}{\sqrt{b}} \, {}_2 F_1 \left( \frac{1}{2}, \frac{1}{2}; 1 ; \frac{b-a}{b} \right)
& \left( \left| {\rm arg} \left( \frac{a}{b} \right) \right| < \pi  \right).
\end{cases}
\end{align*}
As a consequence, we obtain $S_1 = \frac{2}{\pi} \frac{1}{\sqrt{- \eta_4^{\, +}}} \, {}_2 F_1 \left( \frac{1}{2}, \frac{1}{2}; 1; \frac{\eta_4^{\, +} - \eta_4^{\, -}}{\eta_4^{\, +}} \right)$.
Notice that $S_1$ is invariant under $\eta_4^{\, +} \leftrightarrow \eta_4^{\, -}$.
In a similar way, $S_2$ and $S_3$ in Table~\ref{t:class} can be calculated, with the result summarized in Table~\ref{t:G4}.

At the end of this subsection, we discuss the validity of the calculation of $G$ in Table~\ref{t:G4}.
One of the necessary conditions of the validity is that $G$ reduces to the relation~(\ref{eq:G3}) in the limit of $\theta_4 \rightarrow 0$.
Another necessary condition is that $G$ should be invariant under $\eta_4^{\, +} \leftrightarrow \eta_4^{\, -}$,
which corresponds to the invariance of $G$ under $\theta_i \rightarrow - \theta_i$ ($i=1,2,3,4$).
Both conditions are satisfied by the $G$ in Table~\ref{t:G4}.

\begin{table}[bth]
\caption{Calculation of $G = G(\cos \theta_1, \ldots, \cos \theta_4)$, which is invariant under $\eta_4^{\, +} \leftrightarrow \eta_4^{\, -}$.}
\centering
\begin{tabular}{cl}
\hline\noalign{\smallskip}
$(\eta_4^{\, +}, \eta_4^{\, -} )$ & $G(\cos \theta_1, \ldots, \cos \theta_4)$   \\
\noalign{\smallskip}\hline\noalign{\smallskip}
$(+, +)$ & $0$ \\
$(-, -)$ & $\frac{2}{\pi} \frac{1}{\sqrt{- \eta_4^{\, +}}} \, {}_2 F_1 \left( \frac{1}{2}, \frac{1}{2}; 1; \frac{\eta_4^{\, +} - \eta_4^{\, -}}{\eta_4^{\, +}} \right)$ \\
$(+, -)$ & $\frac{2}{\pi} \frac{1}{\sqrt{+ \eta_4^{\, +}}} \, {}_2 F_1 \left( \frac{1}{2}, \frac{1}{2}; 1; \frac{\eta_4^{\, -}}{\eta_4^{\, +}} \right)$ \\
$(-, +)$ & $\frac{2}{\pi} \frac{1}{\sqrt{+ \eta_4^{\, -}}} \, {}_2 F_1 \left( \frac{1}{2}, \frac{1}{2}; 1; \frac{\eta_4^{\, +}}{\eta_4^{\, -}} \right)$ \\
\noalign{\smallskip}\hline
\end{tabular}
\label{t:G4}
\end{table}

\section{Generalization to Jacobi function}
\label{sec:jacobi}

So far, we have dealt with the Legendre function.
In this section, we examine whether or not a relation analogous to (\ref{eq:f_nu}) holds for the Jacobi function $P_\nu^{(\alpha, \beta)}$.
While the analogous relation has been known for one $P_\nu^{(\alpha, \beta)}$, it has been not for the product of $P_\nu^{(\alpha, \beta)}$'s.
However, it is found that the analogous relation holds for $P_\nu^{(\alpha, \beta)} (x) \, P_\nu^{(\beta, \alpha)} (y)$.
(Notice that the superscripts $\alpha, \beta$ are exchanged in order.)
Denote by $\hat{P}_\nu^{(\alpha, \beta)} (x)$ the unnormalized Jacobi function as
$P_\nu^{(\alpha, \beta)} (x) = \frac{\Gamma (\nu+\alpha +1)}{\Gamma (\alpha + 1) \Gamma (\nu+1)} \, \hat{P}_\nu^{(\alpha, \beta)} (x)$, where
\begin{align*}
\hat{P}_\nu^{(\alpha, \beta)} (x) = {}_2 F_1 \left( \nu + \alpha + \beta +1, - \nu; \alpha + 1; \frac{1-x}{2} \right) \qquad (\alpha, \beta > -1).
\end{align*}

\begin{proposition}

For one or two Jacobi function(s), we have the following formula $(2 \gamma := \alpha + \beta + 1)$:
\begin{align}
f_\nu = \sum_{n=0}^\infty c_{\nu,n}^{(\gamma)} \, f_n, \qquad f_\nu =
\begin{cases}
N_\nu^{(\gamma)} \, \hat{P}_\nu^{(\alpha, \beta)} (x) &  ( x \in D_1), \\
N_\nu^{(\gamma)} \, \hat{P}_\nu^{(\alpha, \beta)} (x) \, \hat{P}_\nu^{(\beta, \alpha)} (y) & ( (x,y) \in D_2),
\end{cases}
\label{eq:jacobi}
\end{align}
where $c_{\nu, n}^{(\gamma)} = \frac{\sin \pi (\nu-n)}{\pi} \left( \frac{1}{\nu-n} - \frac{1}{\nu + n + 2 \gamma} \right)$,
$N_\nu^{(\gamma)} = \frac{\Gamma (\nu + 2 \gamma)}{\Gamma (\nu+1)}$, and $D_n$ (generalization of $D_1, D_2$, for later convenience) is given by
\begin{align*}
D_n = \Big \{ ( \cos \theta_1, \ldots, \cos \theta_n ) \in [-1, \, 1]^n  \, \Big \arrowvert \, \sum_{i=1}^n | \theta_i | < \pi, \; \theta_i \in \mathbb R \; (i=1,\ldots, n) \Big \}.
\end{align*}

\label{prop:jacobi}
\end{proposition}

\begin{remark}
For $2 \gamma = 1,2,\ldots$, the sum over $n \in \mathbb N$ in (\ref{eq:jacobi}) can be replaced as
\begin{align*}
\sum_{n=0}^\infty c_{\nu,n}^{(\gamma)} \, f_n \longrightarrow \sum_{n=-\infty}^\infty {\rm sinc} \, (\nu - n) \, f_n \qquad (2 \gamma = 1,2,\ldots).
\end{align*}

\label{rem:sample}
\end{remark}

\begin{remark}
For the case of $f_\nu \propto \hat{P}^{(\alpha, \beta)} (x) \, \hat{P}_\nu^{(\beta, \alpha)} (y)$, the upper subscripts $\alpha, \beta$ are exchanged.
This is due to the relation
\begin{align*}
P_n^{(\alpha, \beta)} (-x) = (-1)^n \, P_n^{(\beta, \alpha)} (x)  \qquad (\forall n \in \mathbb N).
\end{align*}

\label{rem:parity}
\end{remark}

\begin{remark}

Considering that we have the relations
\begin{align*}
D_1 = \{ x \in \mathbb [-1, \, 1] \, | \, (x,y) \in D_2, \; y=1 \}, \qquad \hat{P}_\nu^{(\beta, \alpha)} (1) = 1 \quad (\forall \nu \in \mathbb C),
\end{align*}
we find that in the limit of $y \rightarrow 1$,
the case of $f_\nu \propto \hat{P}_\nu^{(\alpha, \beta)} (x) \, \hat{P}_\nu^{(\beta, \alpha)} (y)$ reduces to the case of $f_\nu \propto \hat{P}_\nu^{(\alpha, \beta)} (x)$

\label{rem:reduction}
\end{remark}

\noindent
{\it Proof}.
As mentioned in Remark~\ref{rem:reduction}, it is sufficient to prove the case where $f_\nu$ is composed of two Jacobi functions.
%
Let $g_x = g_x (y) := f_\nu  - \sum_{n=0}^\infty c_{\nu,n}^{(\gamma)} \, f_n$,
where $f_\nu = N_\nu^{(\gamma)} \, \hat{P}_\nu^{(\alpha, \beta)} (x) \, \hat{P}_\nu^{(\beta, \alpha)} (y)$.
Then the condition of $g_x = 0$ can be rewritten as
\begin{align*}
g_x = 0 \Longleftrightarrow
\begin{cases}
\mathcal H^{(\beta, \alpha)}_\nu g_x = 0, \\
g_x (1) = 0, \\
g'_x (1) = 0,
\end{cases}
\end{align*}
where $\mathcal H^{(\beta, \alpha)}_\nu$, 
generalization of ${\mathcal H}_\nu \; (= \mathcal H^{(0,0)}_\nu)$ in Sec.~\ref{sec:legendre}, is given by
\begin{align*}
\textstyle
\left( \mathcal H^{(\beta, \alpha)}_\nu \varphi \right ) (x) = \left[ (1-x^2) \frac{\partial^2}{\partial x^2} + [ \alpha - \beta - (2 \gamma + 1) x ] \frac{\partial}{\partial x} +  \nu (\nu + 2 \gamma) \right] \varphi (x),
\end{align*}
and the prime $(')$ in $g'_x$ represents the derivative with respect to $y$.
Considering that $\mathcal H^{(\beta, \alpha)}_\nu \, \hat{P}_\nu^{(\beta, \alpha)} = 0$, $\hat{P}_\nu^{(\beta, \alpha)} (1) = 1$,
and $\left. \frac{\rm d}{{\rm d} y} \hat{P}_\nu^{(\beta, \alpha)} (y) \right |_{y=1} = \frac{\nu (\nu+ 2 \gamma)}{2 (1+ \beta)}$,
we obtain
\begin{align*}
\begin{cases}
\mathcal H^{(\beta, \alpha)}_\nu g_x = 0 & \Longleftrightarrow \quad \sum_{n=0}^\infty (-1)^n \, (2n + 2 \gamma) \, f_n = 0, \\
g_x (1) = 0 & \Longleftrightarrow \quad h= 0, \\
g'_x (1) = 1 & \Longleftrightarrow \quad \mathcal H^{(\alpha, \beta)}_0  h = 0,
\end{cases}
\end{align*}
where $h = h(x) := g_x (1)$ and $\mathcal H^{(\alpha, \beta)}_0 = \mathcal H^{(\beta, \alpha)}_{\nu=0}$.
Recall that $h \in L^2 ( [-1, 1] )$ is vanishing for $-1 < x \leq 1$~\cite{jacobi2,jacobi1}, so is $\mathcal H^{(\alpha, \beta)}_0 h$.
Thus to derive $g_x = 0$, the remaining thing is to show the relation $\sum_{n=0}^\infty (-1)^n (2n + 2 \gamma ) \, f_n = 0$ for $(x,y) \in D_2$.
To show it, a convenient way is to use the completeness relation for $\{ \hat{P}_n^{(\alpha, \beta)} \}_{n \in \mathbb N}$, given by
\begin{align*}
\frac{1}{2^{2 \gamma} \Gamma (\alpha +1) \Gamma (\beta + 1)} \sqrt{w (x)} \sqrt{w (-y)} \sum_{n=0}^\infty (-1)^n \, (2n+2\gamma) \, f_n = \delta (x+y),
\end{align*}
where $w (x) = (1-x)^\alpha (1+x)^\beta$ and use has been made of the relation in Remark~\ref{rem:parity}.
Noticing that $D_2 = \{ (x,y) \in \mathbb R^2 \, | \, x+ y > 0, \; -1 < x,y \leq 1 \}$, we find that
\begin{align*}
\sum_{n=0}^\infty (-1)^n \, (2n + 2 \gamma) \, f_n = 0 \qquad ( (x,y) \in D_2 ).
\end{align*}
As a consequence, it is shown that $g_x = 0$ for $(x,y) \in D_2$. This ends the proof.
\hfill{$\Box$}

In the latter half of this section, we examine the case where $f_\nu$ in Prop.~\ref{prop:jacobi} is given by the product of more than two Jacobi functions.
For the case where
\begin{align}
f_\nu = N_\nu^{(\gamma)} \prod_{i=1}^N \hat{P}_\nu^{(\alpha_i, \beta_i)} (x_i), \qquad (x_1,x_2, \ldots, x_N) \in D_N,
\label{eq:f_nu=}
\end{align}
it is found that for $N \geq 3$, the relation of $\alpha_i = \beta_i \; (= \gamma - \frac{1}{2} )$ (for $i=1,2,\ldots, N$) is necessary and sufficient in order that
$f_\nu$ satisfies the same relation as Prop.~\ref{prop:jacobi}.
First we show the necessity. To begin with, we consider the case of $N=3$.
Recall that $\hat{P}_\nu^{(\alpha, \beta)} (1) = 1$, and that in the limit of $x_i \rightarrow 1$, it follows that $(x_j, x_k) \in D_2$ for
$\{ i, j, k \} = \{ 1, 2, 3 \}$ with $(x_1, x_2, x_3 ) \in D_3$. That is, in the limit of $x_i \rightarrow 1$,
the relation $f_\nu = \sum_{n=0}^\infty c_{\nu,n}^{(\gamma)} \, f_n$ for $N=3$ turns out to be that for $N=2$, so that it is required that
\begin{align*}
\alpha_j = \beta_k, \qquad \beta_j = \alpha_k \qquad (j,k = 1,2,3; \, j \neq k),
\end{align*}
from which we obtain $\alpha_i = \beta_i \; (= \gamma - \frac{1}{2} )$ (for $i=1,2,3$).
For $N \geq 4$, we obtain in a similar way the same relation as $\alpha_i = \beta_i \; ( = \gamma - \frac{1}{2} )$ for $i=1,2,\ldots, N$.

The next thing is to show the sufficiency.
For $\alpha = \beta \; (= \gamma - \frac{1}{2})$,
$\hat{P}_\nu^{(\alpha, \beta)} (x)$ can be written as
the following integral representation~\cite{gradshteyn}:
\begin{align*}
\hat{C}_\nu^\gamma (\cos \theta)
= \frac{ 2^\gamma } {B (\frac{1}{2}, \gamma) }  \frac{ 1 }{\sin^{2 \gamma -1} \theta} \int_0^\theta
\frac{ \cos \left[ (\nu + \gamma) \psi \right] }{ ( \cos \psi - \cos \theta)^{1 - \gamma }  } \, {\rm d} \psi \quad ({\rm Re} \, (\gamma) > 0, \; 0 < \theta < \pi),
\end{align*}
where $\hat{C}_\nu^\gamma (x) = \hat{P}_\nu^{(\gamma - \frac{1}{2}, \gamma - \frac{1}{2})} (x)$, unnormalized Gegenbauer function;
and $B (a, b) = \frac{\Gamma (a) \Gamma (b)}{\Gamma (a+b)}$,
Beta function.
Then it is found from the sampling theorem that
for $2 \gamma \in \{ 1,2, \ldots \}$,
$f_\nu := N_\nu^{(\gamma)} \, \prod_{i=1}^N \hat{C}_\nu^\gamma (x_i)$ [for $(x_1,\ldots, x_N) \in D_N$] satisfies the relation
\begin{align}
f_\nu & = \sum_{n=-\infty}^\infty {\rm sinc} \, (\nu - n) \, f_n \notag \\ 
& = \sum_{n=0}^\infty c_{\nu,n}^{(\gamma)} \, f_n,
\label{eq:f=gegen}
\end{align}
where in the second equality, use has been made of the relation in Remark~\ref{rem:sample}.
Although the condition of $2 \gamma \in \{ 1,2,\ldots \}$ is necessary to guarantee the first equality in (\ref{eq:f=gegen}),
this condition on $2 \gamma$ can be relaxed if we rewrite $f_\nu$ as the right-hand side of the second equality.
This is because if $\sum_{n=0}^\infty (-1)^n \, (2n + 2 \gamma) \, f_n = 0$ is satisfied,
then $f_\nu = \sum_{n=0}^\infty c_{\nu,n}^{(\gamma)} \, f_n$ can be derived inductively with respect to $N$, as is shown in the proof of Prop.~\ref{prop:jacobi}.
On the other hand, the relation $\sum_{n=0}^\infty (-1)^n \, (2n + 2 \gamma) \, f_n = 0$ itself
can also be derived inductively with respect to $N$ by using the addition theorem for $\hat{C}_\nu^\gamma (x)$ as~\cite{watson}
\begin{align*}
\hat{C}_n^\gamma (\cos \alpha ) \, \hat{C}_n^\gamma (\cos \beta)
= \frac{1}{B (\frac{1}{2}, \gamma)} \int_0^\pi \sin^{2 \gamma - 1} \omega \cdot \, \hat{C}_n^\gamma (\cos \Omega) \, {\rm d} \omega \qquad ({\rm Re} \, (\gamma) > 0),
\end{align*}
as was performed in the calculation of $G$ in Sec.~\ref{sec:legendre}. 

To summarize,
the allowed values of $\alpha_i, \beta_i$ in order for $f_\nu$ in (\ref{eq:f_nu=}) to satisfy the relation in Prop.~\ref{prop:jacobi} are given in Table~\ref{t:alpha_beta},
where $\gamma$ can be chosen as arbitrary, at least, for ${\rm Re} \, (\gamma)  > 0$.
(This condition may be further relaxed, as long as $f_\nu$ is analytically continued with respect to $\gamma$.)

\begin{table}[bth]
\caption{Condition for $\alpha_i$ and $\beta_i$ $(i= 1,2,\ldots N)$ so that $f_\nu$ in (\ref{eq:f_nu=}) satisfies the relation in Prop.~\ref{prop:jacobi}.}
\centering
\begin{tabular}{ll}
\hline\noalign{\smallskip}
$N$ & $(\alpha_i, \beta_i)$ \\
\noalign{\smallskip}\hline\noalign{\smallskip}
$2$ & $(\alpha_1, \beta_1) = (\beta_2, \alpha_2)$  \\
$3,4,\ldots$ & $\alpha_i = \beta_i \; (= \gamma - \frac{1}{2}) \quad (i=1,2,\ldots,N)$ \\
\noalign{\smallskip}\hline
\end{tabular}
\label{t:alpha_beta}
\end{table}

\section{Conclusion}
\label{sec:conclusion}

We have found that in the case of $\alpha_i = \beta_i \; (= \gamma - \frac{1}{2})$ for $i=1,2,\ldots N$,
$f_\nu$ in (\ref{eq:f_nu=}) satisfies the second equality in (\ref{eq:f=gegen}) for any number $N \in \mathbb N$.
For the case of $\alpha_i \neq \beta_i$, however, the same relation holds for $N=1,2$ only.
The point is that in the case of two Jacobi functions, $f_\nu$ in (\ref{eq:jacobi}) is given by being proportional to
$P_\nu^{(\alpha, \beta)} (x) \, P_\nu^{(\beta, \alpha)} (y)$, not $P_\nu^{(\alpha, \beta)} (x) \, P_\nu^{(\alpha, \beta)} (y)$,
as emphasized in Remark~\ref{rem:parity} and further discussed in the latter half of Sec.~\ref{sec:jacobi}.

An analogous relation holds for other functions, such as the Hermite function $H_\nu (x)$.
Considering that $H_\nu (x)$ is given by
$H_\nu (x) = 2^\nu \lim_{\gamma \rightarrow \infty} \gamma^{\frac{\nu}{2}} \, \hat{C}_\nu^\gamma \left( \frac{x}{\sqrt{\gamma}} \right)$,
we obtain~\cite{kuwata,formula}
\begin{align}
f_\nu = \sum_{n=0}^\infty {\rm sinc} \, (\nu - n) \, f_n, \quad f_\nu =
\begin{cases}
\frac{1}{2^{2 \nu}} \frac{1}{\Gamma (\nu+1)} \, H_{2 \nu + \varepsilon} (x) & (x > 0; \; \varepsilon = 0,1), \\
\frac{1}{2^\nu} \frac{1}{\Gamma (\nu+1)} \, H_\nu (x) \, H_\nu (y) & (x + y > 0).
\end{cases}
\label{eq:hermite}
\end{align}
It is tempting to ask whether or not there exists an analogous relation for more than two $H_\nu$'s.
Consider the case where $f_\nu$ is given by the product of multiple $H_\nu$'s as
$f_\nu \propto \prod_{n=1}^N H_{k \nu + \varepsilon} (x_i)$ (for $k \in \{ 1,2,\ldots \}, \varepsilon \in \{ 0,1,\ldots, k-1 \}$).
In this case,
the applicability of the sampling theorem to the corresponding Gegenbauer function
$\hat{C}_{k \nu + \varepsilon}^\gamma (x)$ implies that the relation $\frac{1}{2} k N = 1$ is required, 
so that for $k, N  \in \{ 1,2,\ldots \}$, it follows that $(k, N) = (2,1), (1,2)$, as is realized in eq.~(\ref{eq:hermite}).
The allowed values of $N$, together with the known values, for various functions are summarized in Table~\ref{t:summary},
where the Legendre function is a special case of the Gegenbauer function.

\begin{table}[bth]
\caption{Allowed and known values of $N$ such that the sampling theorem based summation formula holds for the product of $N$ jacobi, Gegenbauer, and Hermite functions.}
\label{tab:summary}       
\centering
\begin{tabular}{lll}
\hline\noalign{\smallskip}
Function & Allowed $N$ & Known $N$   \\
\noalign{\smallskip}\hline\noalign{\smallskip}
Jacobi & $1,2$  & $1$ \\
Gegenbauer  & $1,2,3,\ldots$  & $1,2$  \\
Hermite & $1, 2$ & $1,2$  \\
\noalign{\smallskip}\hline
\end{tabular}
\label{t:summary}
\end{table}

\section*{Acknowledgements}
The authors are indebted to T. Kobayashi and H. Fujisaka for stimulating discussions and suggestions.  

\end{document}